\documentclass[final,5p,times,twocolumn]{elsarticle}
\usepackage{amssymb}
\usepackage{lineno}
\usepackage{wasysym}
\usepackage{color}
\usepackage{epstopdf}

\usepackage{mathrsfs}
\usepackage{siunitx}
\usepackage[latin1]{inputenc}
\usepackage{bbm}
\usepackage{graphicx}
\usepackage{graphics}
\usepackage{epsfig}
\usepackage{xcolor}
\usepackage{float}
\usepackage{ulem}

\journal{Physics Letters A}

\begin{document}

\begin{frontmatter}

\title{Kinematic Vortices induced by defects in Gapless Superconductors}

\author{{V.S. Souto$^{1}$, E.C.S Duarte$^{1}$}, E. Sardella$^{2}$, R. Zadorosny\corref{*}$^{1}$}
\ead{rafael.zadorosny@unesp.br}
\cortext[*]{Corresponding author}

\address{$^{1}$Departamento de F\'isica e Qu\'imica, , Universidade Estadual Paulista (UNESP),Faculdade de Engenharia, Caixa Postal 31, 15385-000, Ilha Solteira, SP, Brazil}
\address{$^{2}$Departamento de F\'isica,  Universidade Estadual Paulista (UNESP), Faculdade de Ci\^encias de Bauru,  Caixa Postal 473, 17033-360, Bauru, SP, Brazil}

\begin{abstract}
	
The generalized time-dependent Ginzburg-Landau (GTDGL) theory was first proposed to describe better gap superconductors and the phenomenon of thermal phase-slips (PSs) in defect-free systems. However, there is a lack of information about studies involving PSs in mesoscopic superconductors with surface defects. Thus, in this work, we simulated samples with two co-linear surface defects consisting of a lower $T_c$ superconductor narrowing the sample in its central part. The non-linear GTDGL equations were solved self-consistently under variable applied currents and by considering both gapless and gap-like superconductors. In such systems, the currents passing by the constriction induce the appearance of kinematic vortices even in the gapless sample. The dynamics always occur with a pair forming at opposite edges of the sample and annihilating in the center. It is noticed that the resistive state appears at distinct values of the applied current density for different samples, and the critical current presents a tiny difference between gapless and gap-like samples. It is worth mentioning that parameters such as the size of electrical contacts and constriction affect the critical current and the average velocity of the kinematic vortices.

\end{abstract}


\end{frontmatter}


\section{Introduction}

Phase-slips (PSs) on superconductors drive them to a non-zero resistivity state\cite{MeyerMinnigerode,Tinkham1974}. Such phenomenon occurs when the depairing current density is locally reached (${J}_{dep}$), breaking the Cooper pairs. For quasi-one-dimensional systems, as nanowires, the PSs are known as phase-slip centers (PSC) consisting of a single point with zero superconducting order parameter \cite{IvlevKopnin1980,IvlevKopnin1984,Goiberg}. On the other hand, for thin superconducting films or micro bridges, the so-called phase-slip lines (PSLs) are formed perpendicularly to the applied current \cite{Yamamoto, Kulikovsky}. In that PSLs, the superconducting order parameter oscillates between zero and its maximum value inducing voltage oscillations in the samples \cite{Beryorovb2009b,Churilov,Galaiko}. Devices like superconducting stripline detectors and single-photon detectors \cite{Miki} can be affected by PSLs since their functioning is based on the appearance of non-zero voltages when a photon or a particle hits them. Besides that, some works showed a dual relationship between PSLs and Josephson junctions, which can be used to develop standard current devices \cite{Mikia, Suzuki, Casaburi, Casaburia, Cristiano, Kyosuke}. Two mechanisms are responsible for the appearance of PSLs, a quantum one \cite{Petkovic} originated from a low free-energy barrier, but not zero, and a thermal-activate mechanism \cite{Petkovic}. This last one can be studied by the generalized time-dependent Ginzburg-Landau (GTDGL) formalism with good agreement with experiments \cite{Petkovic}.

In 1993, by solving the GTDGL equations, Andronov demonstrated that the PSLs are formed by a pair of a vortex and an antivortex moving at very high velocities. Those specimens were known as kinematic vortex (KV), and kinematic antivortex (KAv), respectively \cite{Andronov}.

In 2003, Sivakov and coworkers described a pioneering experiment to measure the velocity of a KV\cite{Sivakov}. They used a tape consisting of a superconductor-normal-superconductor (S/N/S) configuration, with S being Sn. A transport current was applied to the sample, and a velocity of $10^5$ m/s was obtained for the KV.
In addition, within the GTDGL formalism, Berdiyorov et al. \cite{Beryorovb2009b} studied the dynamics of KV in superconducting stripes under the influence of externally applied fields. It was shown that for narrow electrical contacts (with sizes smaller than the width of the samples), the pairs of KV and KAv nucleate at different regions of the sample, and such dynamics are current-dependent. Then, distinct resistive states take place. Besides that, the velocity of KV and KAv is inversely proportional to the external magnetic field, and KVs could become Abrikosov-like ones. Additionally, in Ref. \cite{Berdiyorov2009a}, the authors showed that, for $H =0$, an increasing $\gamma$ (a parameter related with the relaxation of the superconducting order parameter) broaden the range of currents for which a resistive state takes place. However, for gapless superconductors, i.e., $\gamma=0$, the PSLs no longer exist. By applying a magnetic field, the vortex velocity and its dynamics change, and the $I-V$ characteristics become hysteretic, decreasing as the applied magnetic field increases. Vodolazov and coworkers also verified such behavior \cite{Vodolazov2004}. Meanwhile, Presotto et al. shown that a constriction in the middle of a stripe changes the KV dynamics, and external fields can induce surface-barrier effects on those vortices \cite{alice}. 

In Ref.\cite{Berdiyorov2009c}, the authors studied the KV dynamics in superconductors with magnetic dots. They showed that dots magnetized perpendicularly to the plane of the films increase the range of currents for which a resistive state takes place. It was also described that the voltage frequency depends on the number of magnetic dots and their magnetization. 


The material surrounding a superconductor also affects the KV dynamics and its resistive state. Barba-Ortega et al. \cite{Ortega2018}, applying the de Gennes boundary conditions on superconducting stripes, showed that both $J_c$ and the current range of resistive states are dependent on the surrounded material. Meanwhile, a resistive state is avoided when a superconductor with a higher $T_c$ surrounds the sample.

As mentioned before, the phenomenological aspects of PSCs and PSLs are properly described by the GTDGL formalism. However, as pointed out by Kramer and Barotoff, the ordinary time-dependent Ginzburg-Landau (TDGL) equations describe just a narrow range of current densities ($\Delta{J}$) for the resistive state at $\gamma=0$\cite{KramerBaratoff,KramerWattsTobin}. The authors found that a PSL rises from $J_{min}<J<J_{c}$ where $J_{min}=0.325 J_0$ and $J_c=0.335 J_0$, being $J_0$ the depairing current \cite{KramerBaratoff,KramerWattsTobin}. On the other hand, we show that a constriction broadens the range of currents at the resistive state by using the TDGL formalism. Furthermore, significant changes in ${J}_{dep}$ of the simulated systems were verified. To characterize the resistive state, several analyses were carried out for the so-called gap-like and gapless superconductors and the influence of the sizes of metallic contacts and constrictions at the supercurrent distribution and voltage-time characteristics.

\section{Theoretical formalism}

In the present work, the kinematic vortex dynamics were studied applying the GTDGL formalism \cite{Beryorovb2009b,KramerWattsTobin} (equation\ref{eq:gtdgl}). It was considered a thin superconducting film (thickness$<<\lambda$) with a constriction made by two co-linear surface defects, as shown in Figure \ref{fig1}. Such defects consisted of a superconductor with $T_c$ lower than that one of the main matrix. Transport currents were injected into the sample by considering two metallic leads of size $\alpha$.

\begin{eqnarray} \label{eq:gtdgl}
	\frac{u}{\sqrt{1+\gamma^2\left|\psi\right|^2}}\left(\frac{\partial}{\partial t} + i \varphi + \frac{\gamma^2}{2}\frac{\partial\left|\psi\right|^2}{\partial t}\right) \psi = \nonumber \\ \left(\nabla - i \textbf{A}\right)^2\psi + \left(g(\textbf{r}) -{T}- \left|\psi\right|^2\right)\psi
\end{eqnarray}

The equation \ref{eq:gtdgl} is coupled with the scalar potential $\nabla^2\varphi= \mathbf{\nabla} \cdot \mathbf{J}_s = \mathbf{\nabla}\cdot\Im(\psi^\ast(-i\mathbf{\nabla}-{\textbf{A}})\psi)$. The distances are measured in units of the coherence length, $\xi(0)$, temperature in units of the critical temperature, $T_c$, the electrostatic potential by $\varphi_0=\hbar/2e\tau_{GL}(0)$, and time by $\tau_{GL}(0)=\pi\hbar/8k_BT_{c}u$. The parameter $u = 5.79$ is obtained from a microscopic derivation of equation
~\ref{eq:gtdgl} in the dirty limit \cite{KramerWattsTobin}. $\gamma=2\tau_E\psi_0 / \hbar$, with $\tau_E$ being the inelastic electron-collision time. The parameter $g(\textbf{r})=1$ for all regions except in the surface defects (see Fig. \ref{fig1}), where it was set $g(\textbf{r})=0$. Additionally, $T$ was set zero once we were interested in the different behaviours presented by gap-like and gapless systems.

Superconducting-vacuum boundary conditions $\textbf{n}\cdot(\nabla - i \textbf{A})\psi = 0$ and $\textbf{n}\cdot(\nabla\varphi) = 0$ were taken at all boundaries of the sample, except at the interface with the current contacts, where it was used $\psi = 0$ and $\textbf{n}\cdot(\nabla\varphi) = -J$, being $J$ is the external applied current density in units of $J_0 = \frac{c\sigma\hbar}{2e\tau_{GL}(0)}$; $\sigma$ is the normal electrical conductivity.

We numerically solved the equation \ref{eq:gtdgl} by applying the link variable method \cite{Gropp} to preserve the gauge invariance of the discretized equations.

\section{Results and Discussion}

\begin{figure}[!htb]
	\centering
	\includegraphics[width=7cm]{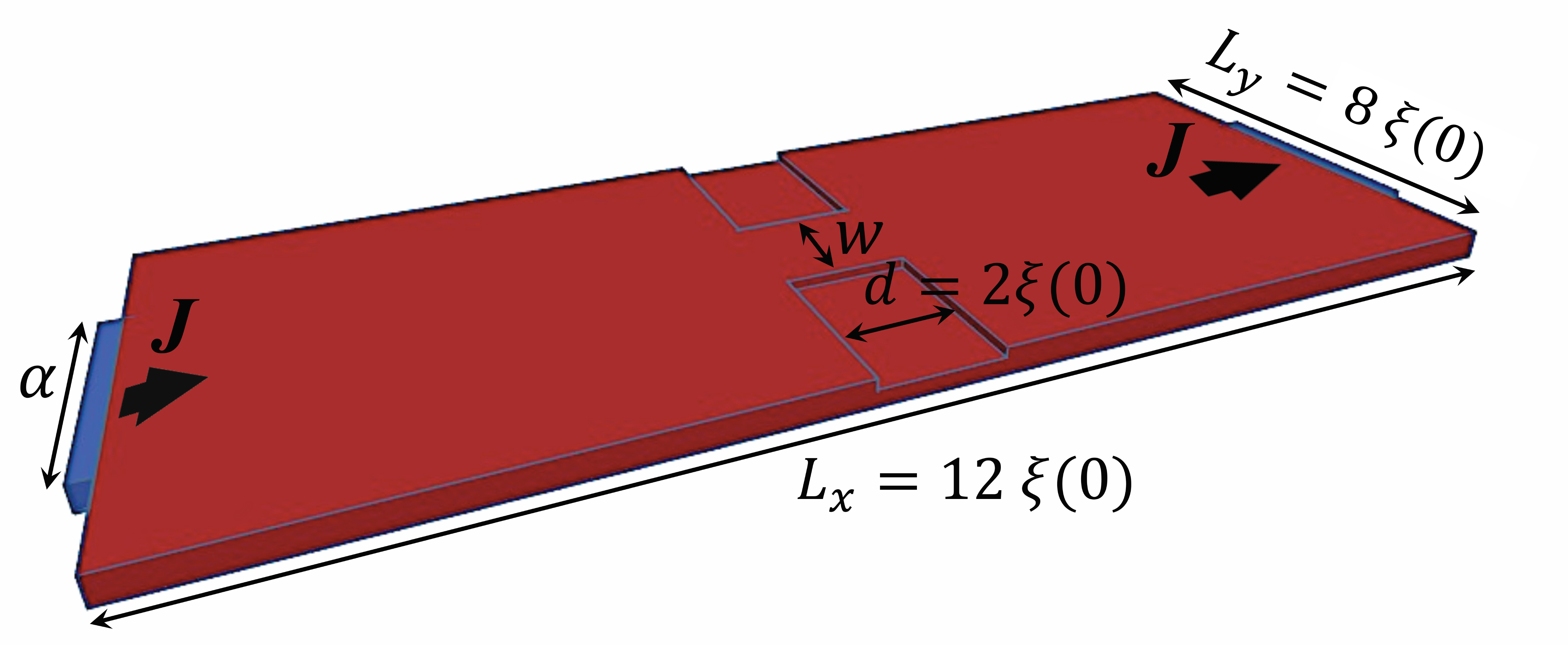}
	\caption{(Color online) It is shown a schematic view of the studied sample. The dark blue of size $\alpha$ corresponds to the attached metallic leads; the dark red area represents the superconducting matrix, and the thinner slits correspond to the lower $T_c$ superconducting defects. $J$ represents the applied transport current, $d=2\xi(0)$ and $w$ are the sizes of the constriction.}
	\label{fig1}
\end{figure}

We studied KV dynamics in two systems. One of them being a gapless superconductor ($\gamma=0$ in equation \ref{eq:gtdgl}), and the other one being a gap-like material ($\gamma=10$ in equation \ref{eq:gtdgl}). The size of the sample was $L_x = 12\xi\left(0\right) \times L_y = 8\xi\left(0\right)$, with a central constriction as shown in Figure\ref{fig1}. The sizes of the electrical leads and the constriction were varied as follows, $\alpha=2\xi\left(0\right)$, $4\xi\left(0\right)$,  $6\xi\left(0\right)$, and $w=2\xi\left(0\right), 4\xi\left(0\right), 6\xi\left(0\right)$, $8\xi\left(0\right)$, respectively. This last value is equivalent for a homogeneous sample.

\subsection{Homogeneous samples}

\begin{figure}[!htb]
	\centering
	\includegraphics[width=7cm]{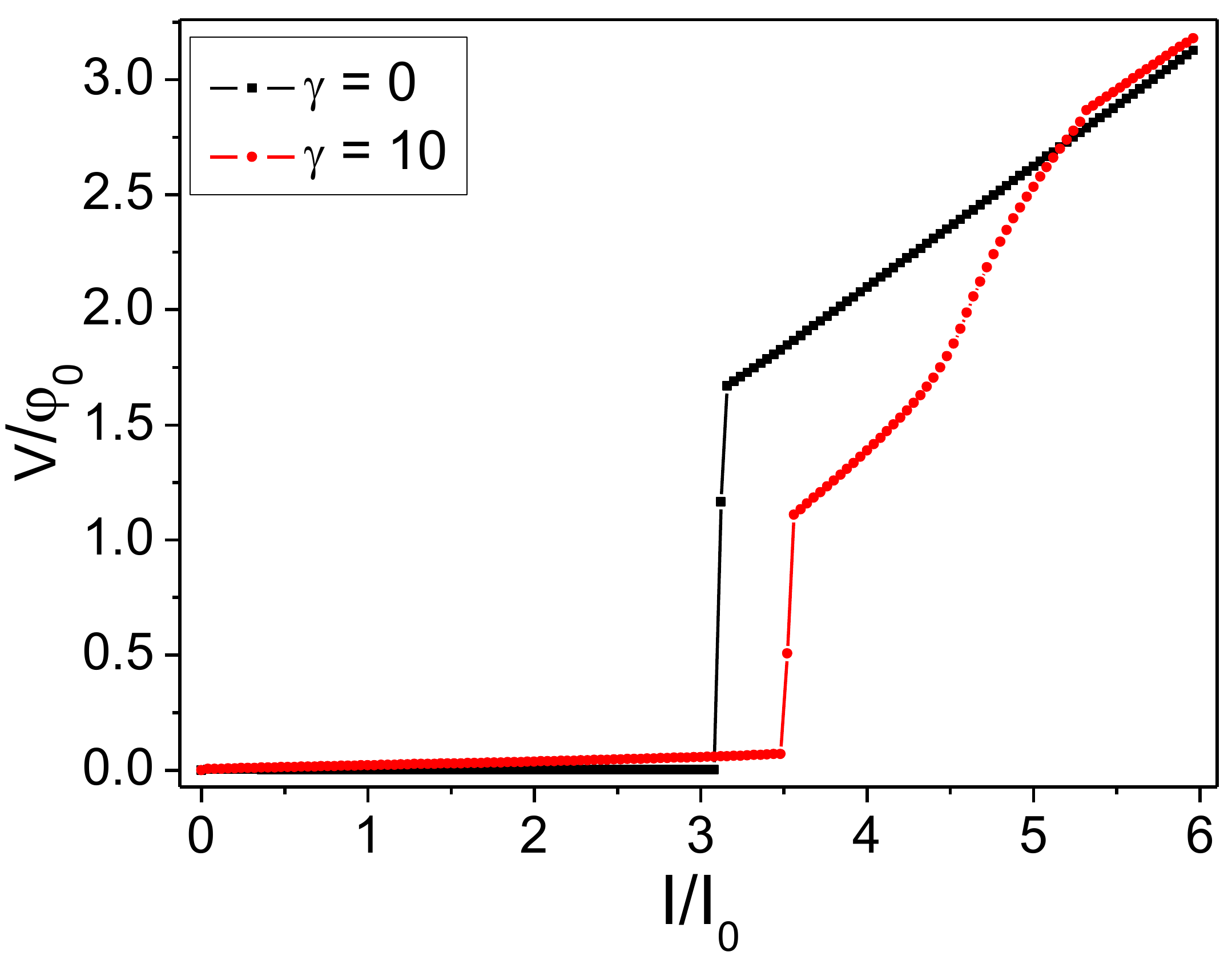}
	\caption{(Color online) $\varphi(J)$ characteristics for the gapless (black line) and gap-like (red line) simulated samples. The former does not present a resistive state.}
	\label{fig2}
\end{figure}

To understand the formation of KVs, the behavior of gap-like and gapless samples for $\alpha=2\xi(0)$ and $w=8\xi\left(0\right)$ was analyzed. As expected, the gapless sample has a sharp transition to the normal state while the gap-like one presents resistive states before the transition to the normal state (see Figure \ref{fig2}), following Ref.\cite{Beryorovb2009b}. 


\begin{figure}[!htb]
	\centering
	\includegraphics[width=8cm]{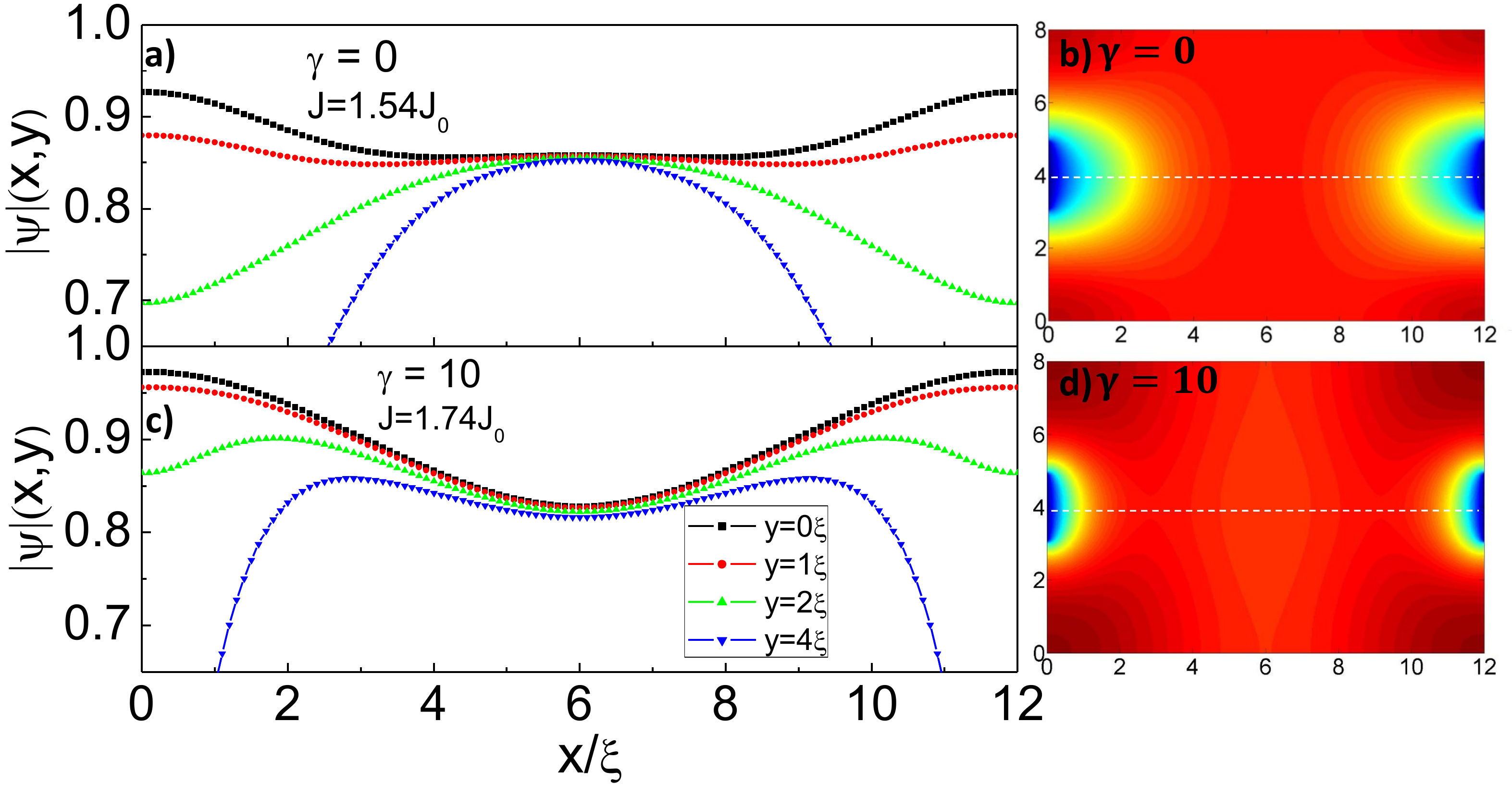}
	\caption{(Color online) $\left|\psi\right|$ as a function of $x$ taken at different positions in $y$-axis (a) for gapless, and (c) for gap-like samples. The snapshots show the distribution of $|\psi|$ at (b) $J(\gamma=0)=1.54 J_0)$, and at (d) $J(\gamma=10)=1.74 J_0)$. The white line in panels (b), and (d) exemplifies where the values of $|\psi|$ were taken to plot the graphs at (a), and (c).}
	\label{fig3}
\end{figure}

Thus, the formation of KV-KAv pairs is, at first, associated with the relaxation of $|\psi|$. However, the distribution of $|\psi|$ presented in Figure \ref{fig3}, shows a local maximum (minimum) at the center of the gapless (gap-like) sample in the $y$-axis.

Nonetheless, if, on the one hand, the behaviors shown in Figures \ref{fig2} and \ref{fig3} are well known, i.e., $\psi$ has a local minimum in the center of the sample for $\gamma=10$; on the other one, it leads us to conclude that KV can be induced in gapless systems by artificially producing a local minimum in $\psi$. Following this, surface defects of a superconductor with lower $T_c$ were considered in the studied systems. 

\subsection{The size-dependence of the metallic leads}

\begin{figure}[!htb]
	\centering
	\includegraphics[width=8cm]{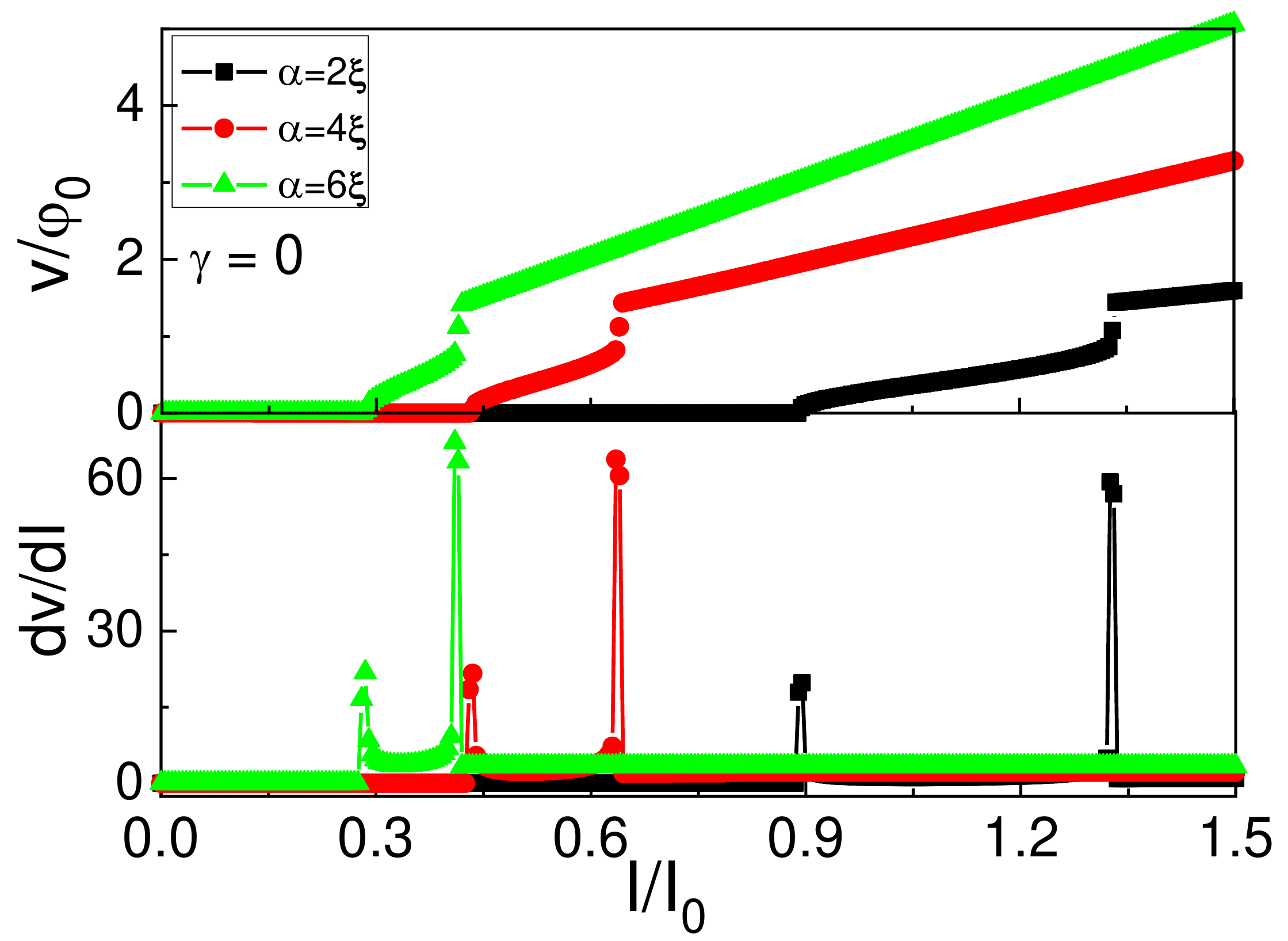}
	\caption{(Color online) $\varphi(J)$ characteristics and differential resistance as a function of applied current (both normalized) for a gapless system. The size of the metallic leads was varied by keeping the constriction sizes at $w=d=2\xi\left(0\right)$. It is noted that $J_{c1}$ decreases as increasing the size of the leads.}
	\label{fig4}
\end{figure}

In Figure \ref{fig4}, the $\varphi(J)$ characteristics and the differential resistance $d\varphi/dJ (J)$ are shown for the gapless system. The simulations were carried out for three different values of $\alpha$, and for $w=d=2\xi\left(0\right)$.

\begin{figure}[!htb]
	\centering
	\includegraphics[width=6.5cm]{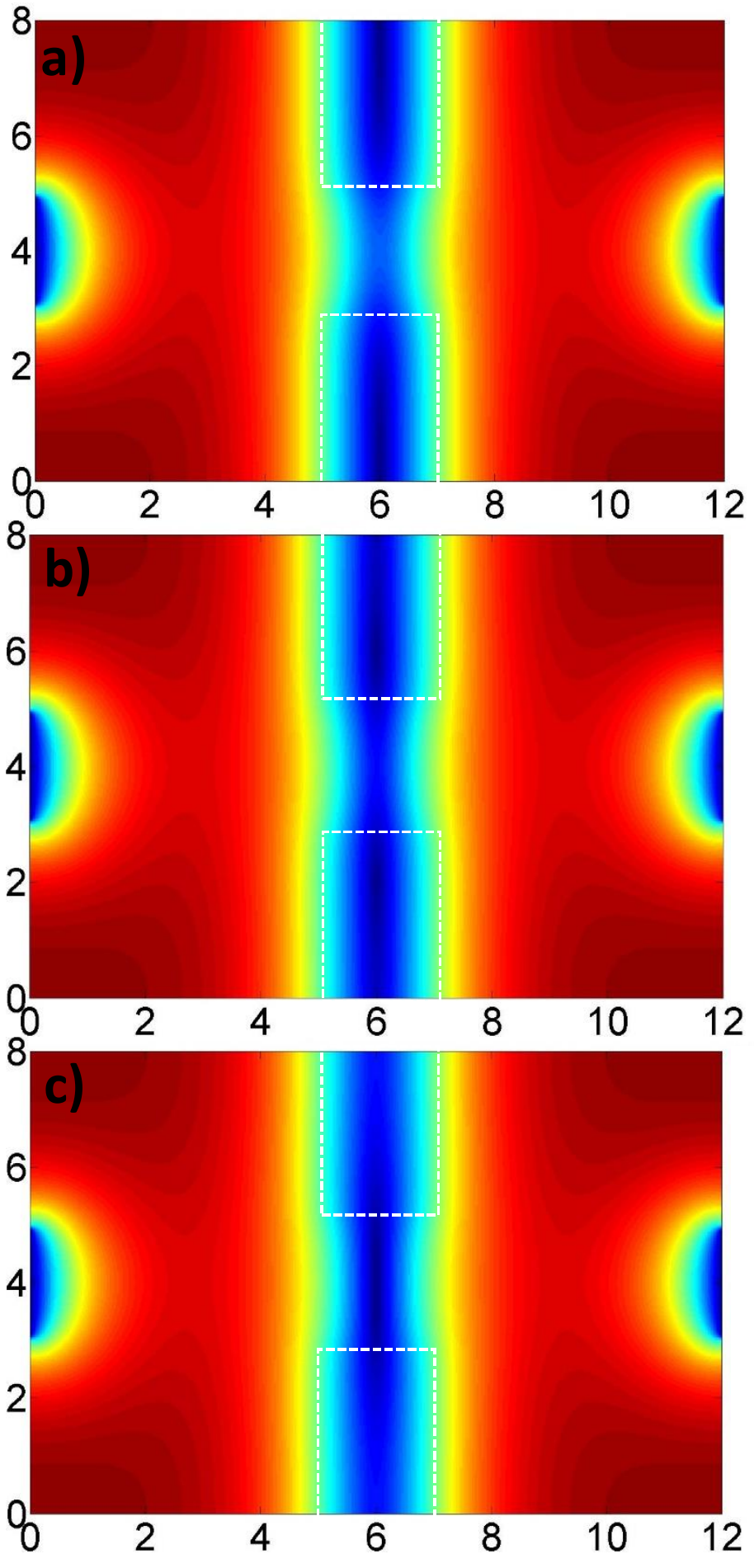}
	\caption{(Color online) Distribution of $\left|\psi\right|$ in a gapless system at $J=0.895J_0$, and metallic leads $\alpha=2\xi\left(0\right)$. The dark red (blue) represents the maximum (minimum) of $|\psi|$.}
	\label{fig5}
\end{figure}

\begin{figure}[!htb]
	\centering
	\includegraphics[width=8cm]{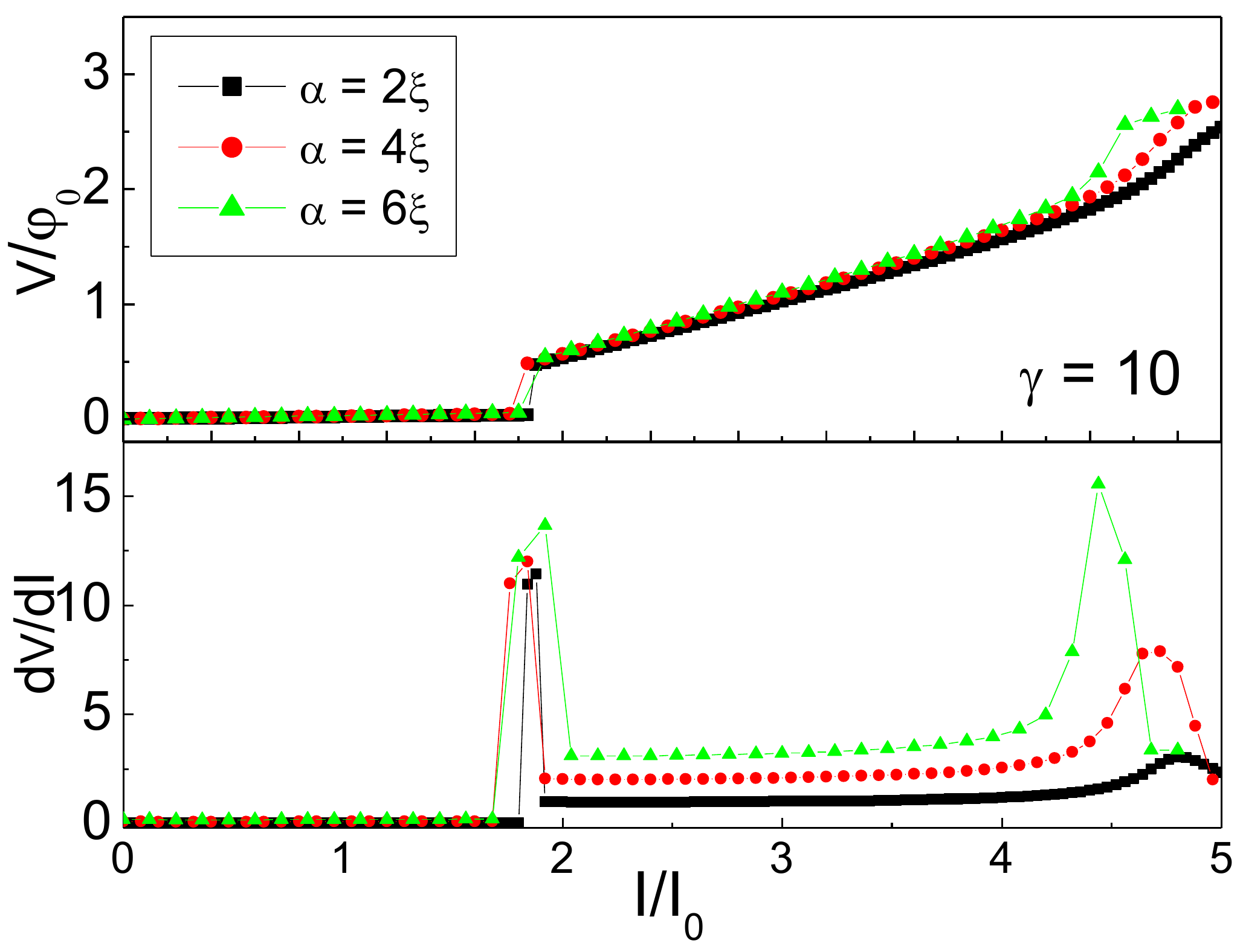}
	\caption{(Color online) $\varphi(J)$ characteristics and differential resistance as a function of the applied current (both normalized) to a gap-like system. The size of the metallic contacts ($\alpha$) was varied by keeping $w=d=2\xi\left(0\right)$. It is noted that $J_{c1}$ is almost constant by varying $\alpha$.}
	\label{fig6}
\end{figure}

Figure \ref{fig4} shows a decreasing of $J_{c1}$ by increasing $\alpha$, with $J_{c1}$ the current of a resistive state beginning. In addition, the range of currents for which the resistive state takes place decreases with $\alpha$. The dynamics are also affected by the constriction, as shown in Figure \ref{fig5}, presenting only one kind of it and being independent of the applied current. This behavior contrasts with that described in Ref. \cite{Beryorovb2009b}. However, the gap-like system acts similarly to the gapless (see Figure \ref{fig6}, which shows only one resistive state), indicating the KV dynamics are sensible to topological patterns in superconducting systems, and in addition, non-resistive states are broadening by using very tiny electrical contacts. In such a case, $J_{c1}$ slightly decreases with the increase of $\alpha$. 

\begin{figure}[!htb]
	\centering
	\includegraphics[width=8.5cm]{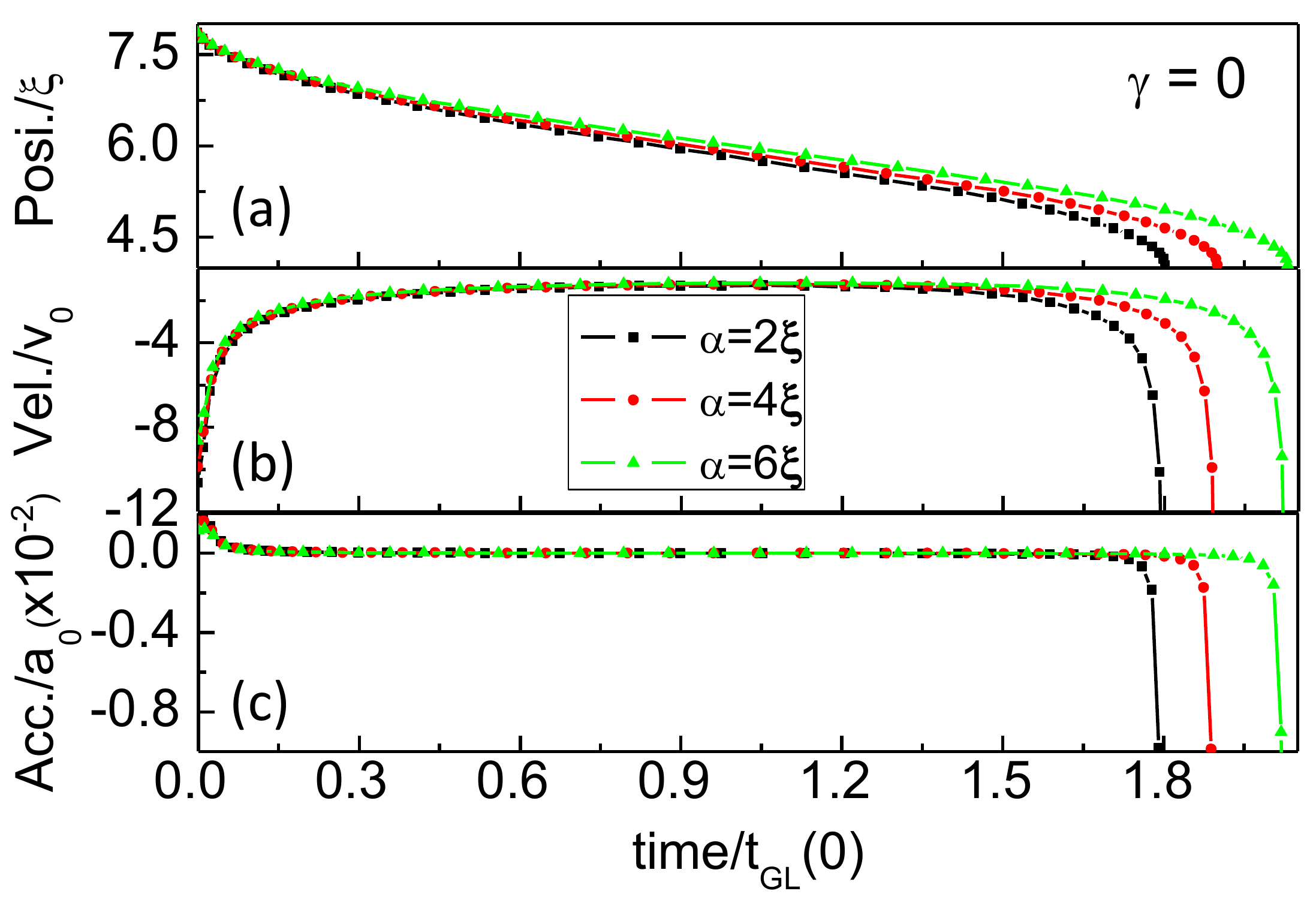} 
	\caption{(Color online) Kinetic curves for the KV. In (a) is shown the position; in (b) the instantaneous velocity, and in (c) the acceleration. All of them plotted as a function of time and obtained at $J_{c1}$. The average velocity is $\alpha$-dependent.}
	\label{fig7}
\end{figure}

Focusing on the gapless sample, Figure \ref{fig7} presents the kinetic data of a KV. All curves were taken at $J=J_{c1}$. As the simulated systems are symmetric, one can analyze only the motion of the KV. In panel (a) is shown the time-dependence of the position. It is noted that the fastest vortices are associated with the smaller leads. The annihilation occur at $1.8t_{GL}$, $1.9t_{GL}$ and $2.0t_{GL}$, for $\alpha=2\xi\left(0\right)$, $\alpha=4\xi\left(0\right)$ and $\alpha=6\xi\left(0\right)$, respectively. In panels (b) and (c), one can also see that the KV is accelerated in its penetration, and a larger acceleration takes place as a consequence of the attractive interaction between KV and KAv.

\subsection{Distribution of superconducting currents in the KV dynamics}

\begin{figure}[!htb]
	\centering
	\includegraphics[width=8.5cm]{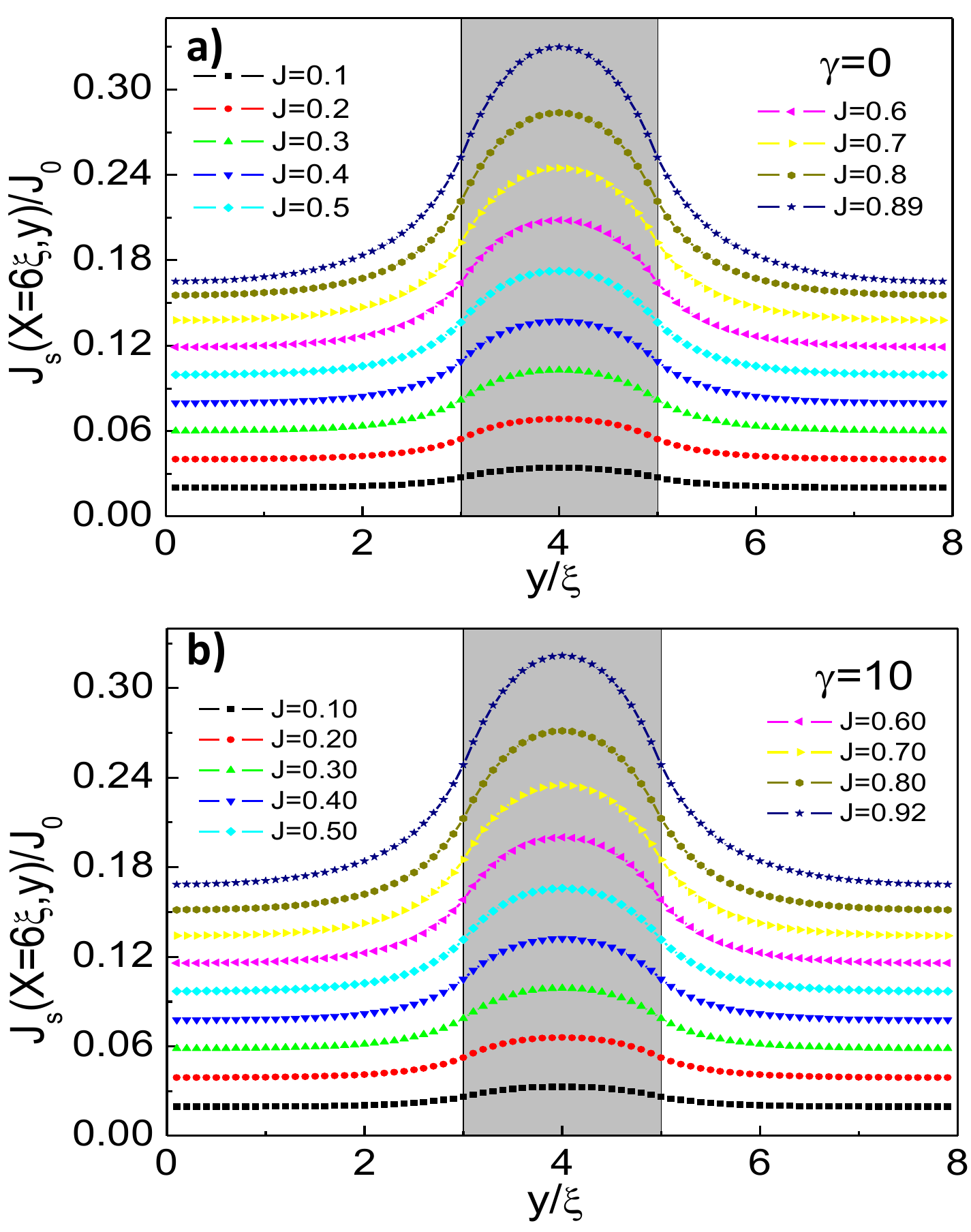}
	\caption{(Color online) Distribution of $J_s$ as a function of $y$ at $x=6\xi\left(0\right)$. (a) gapless and (b) gap-like samples. The curves were taken at zero resistance state, and considering $\alpha=d=w=2\xi\left(0\right)$. The dark area represents the constriction's region.}
	\label{fig8}
\end{figure}

Figure \ref{fig8} shows the evolution of the superconducting current density ($J_s$) before the formation of a KV-KAv pair, for both studied samples, and for $\alpha=d=w=2\xi(0)$. The curves were taken along the middle of the samples (at $x=6\xi(0)$). An increase of the superconducting current density occurs in the constriction, which induces the appearance of a KV-KVa pair in the defects. Both samples present similar behavior.

\begin{figure}[!htb]
	\centering
	\includegraphics[width=6.5cm]{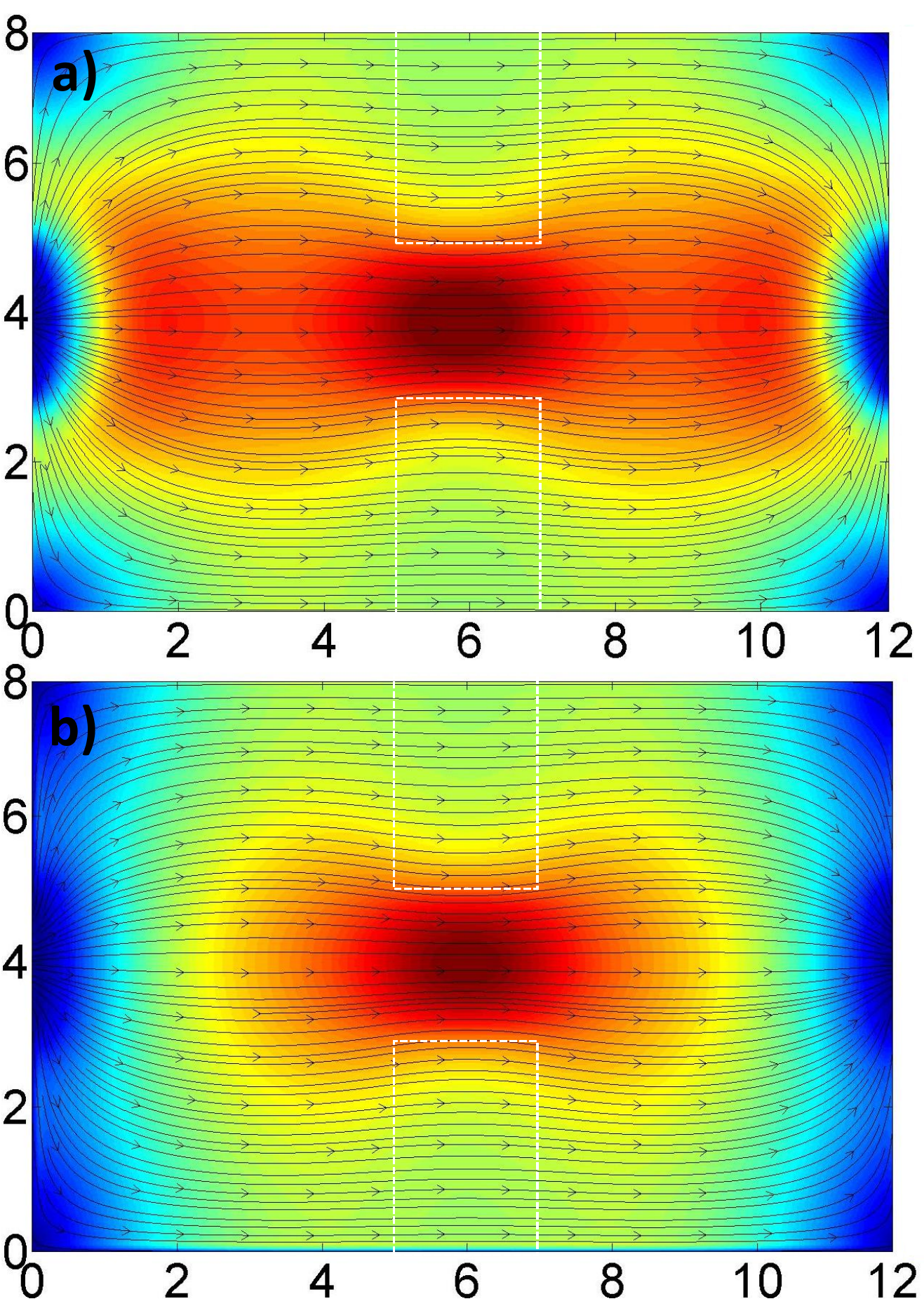}
	\caption{(Color online) Snapshots of $J_s$ for the (a) gapless system at $J=0.89J_0$, and for the (b) gap-like at $J=0.92J_0$. Dark red (blue) indicates the larger (lower) $J_s$. The streamlines indicate the flux of the currents.}
	\label{fig9}
\end{figure}

Figure \ref{fig9} shows snapshots of $J_s$ at $J=0.89J_0$ (for gapless), and at $J=0.92J_0$ (for gap-like). It is evidencing the crowding of the currents in the constriction. The streamlines indicate the $J_s$ flux. It can also be noted that the increase of $J_s$ is restricted to the constriction in the gap-like sample.

\begin{figure}[!htb]
	\centering
	\includegraphics[width=8cm]{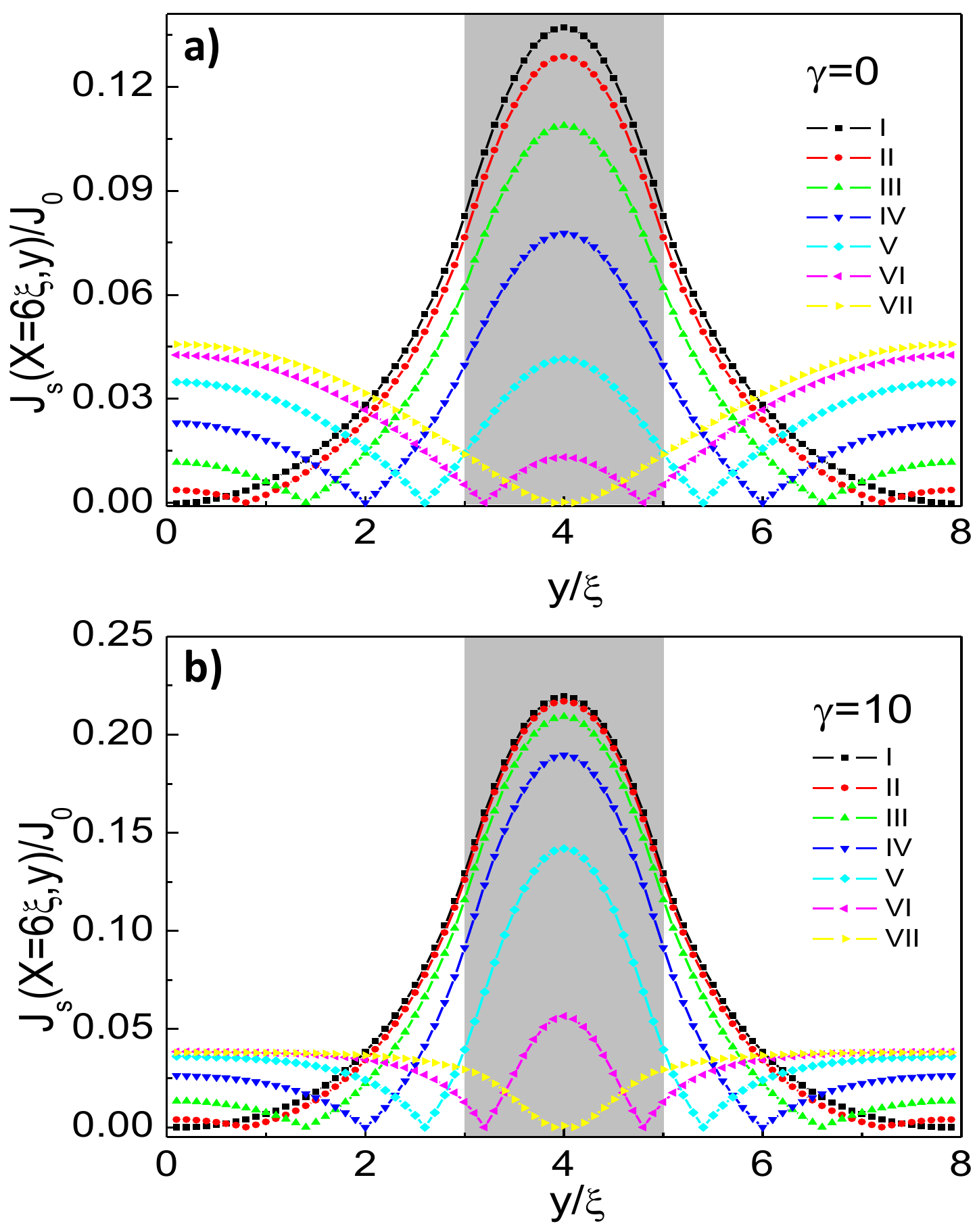}
	\caption{(Color online) $J_s(y)$ at $x=6\xi\left(0\right)$. (a) gapless sample at $J=0.895J_0$ and (b) gap-like at $J=0.925J_0$ (both with $\alpha=w=d=2\xi\left(0\right)$). The numbers from I to VII indicate a temporal evolution without changing the injected current.}
	\label{fig10}
\end{figure}

Figure \ref{fig10} shows, for the same conditions of Figure \ref{fig8}, the $J_s$ distribution during the motion of a KV-KAv pair at $J_{c1}$ of both samples. The time evolution of $J_s$ is shown from I to VII. The pair's position can be identified by the points of zero current, which indicate the vortex core. One can note that the KV (KAv) is formed in the top (bottom) edge of the sample when $J_s$ reaches a maximum value in the center of the constriction. As the KV-KAv pair moves toward the center of the sample, $J_s$ decreases, and it is vanished at the annihilation point (see Figure \ref{fig10} (a) curve VII).

The same analyzes performed for the gapless system was carried out in the gap-like as shown in Figure \ref{fig10} (b). Besides, the general distribution of $J_s$ is similar to that one for the gapless system (Figure \ref{fig10} (a)), there are some particularities related to the relaxation of $\psi$ ($\gamma \neq 0$). First, it is noted a saturation of $J_s$ at the edges of the defects for 0.045$J_0$. As a lower $ T_c$ characterizes the defects, $J_c$ is locally suppressed faster in the case of Figure \ref{fig10} (b) than in (a). This is a consequence of $\gamma \neq 0$ due to a delay in recovering the superconducting state.

\begin{figure}[!htb]
	\centering
	\includegraphics[width=8cm]{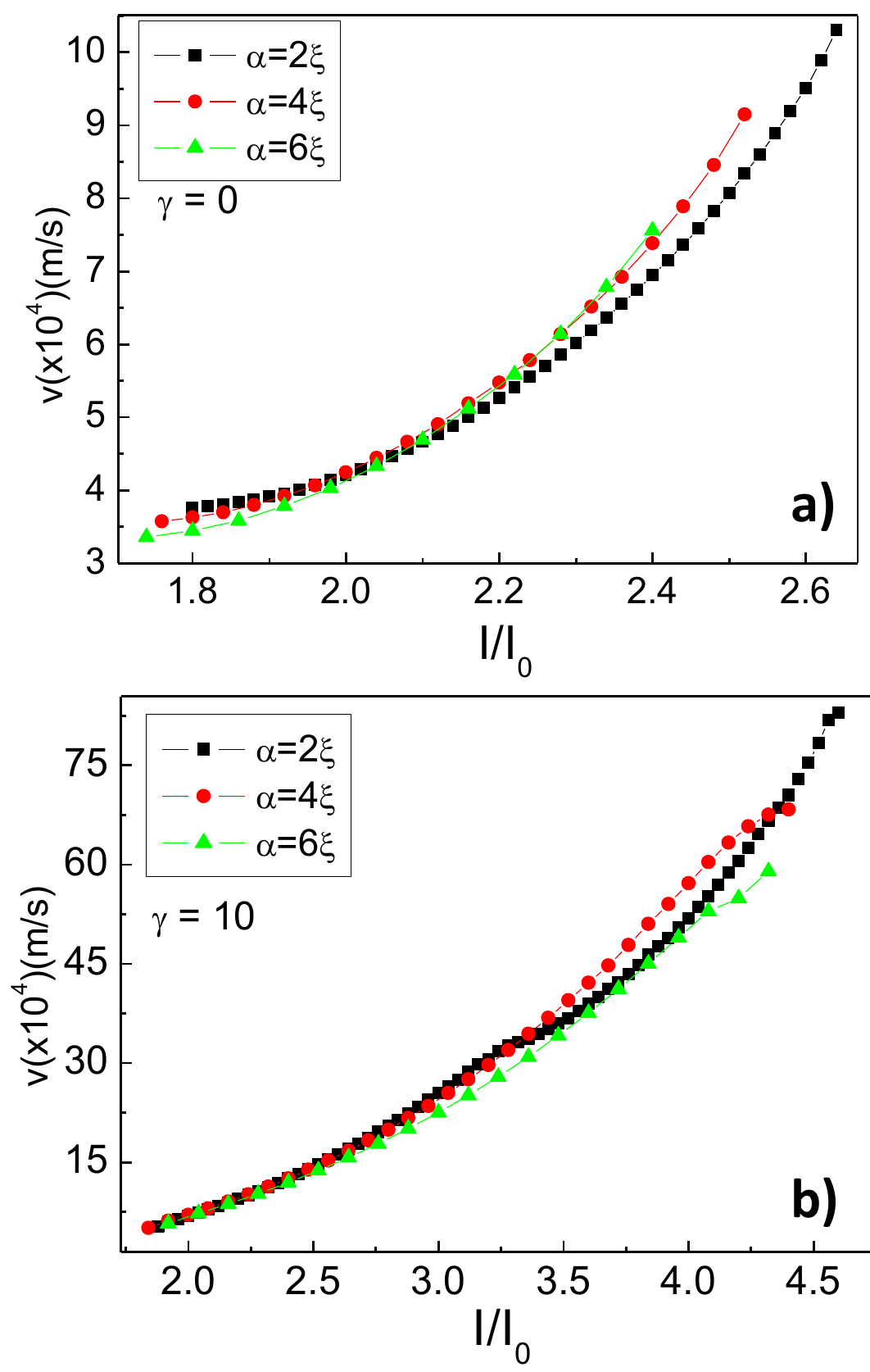}
	\caption{(Color online) Average velocity of (a) gapless and (b) gap-like systems for different sizes of metallic contacts, $\alpha=2,4,6\xi(0)$. }
	\label{fig11}
\end{figure}

The average velocity of the vortex as a function of the applied current in real units (here it was considered the parameters of a PbIn alloy, i.e., $T_c=7.0K$, and $\xi(0)= 30nm$) \cite{Poole}, is shown in Figure \ref{fig11} for both samples. It is seen that the velocity increases faster with $J$ for smaller $\alpha$. The velocities of gap-like samples can be ten times larger than those of gapless systems, and it is related to its larger $J_{c2}$. Such difference in the velocities is also $\gamma$-dependent, leading to a degraded superconducting region and allowing a faster motion of the vortices. 


\subsection{Kinematic vortex dynamics for constrictions with different sizes}

\begin{figure}[!htb]
	\centering
	\includegraphics[width=8cm]{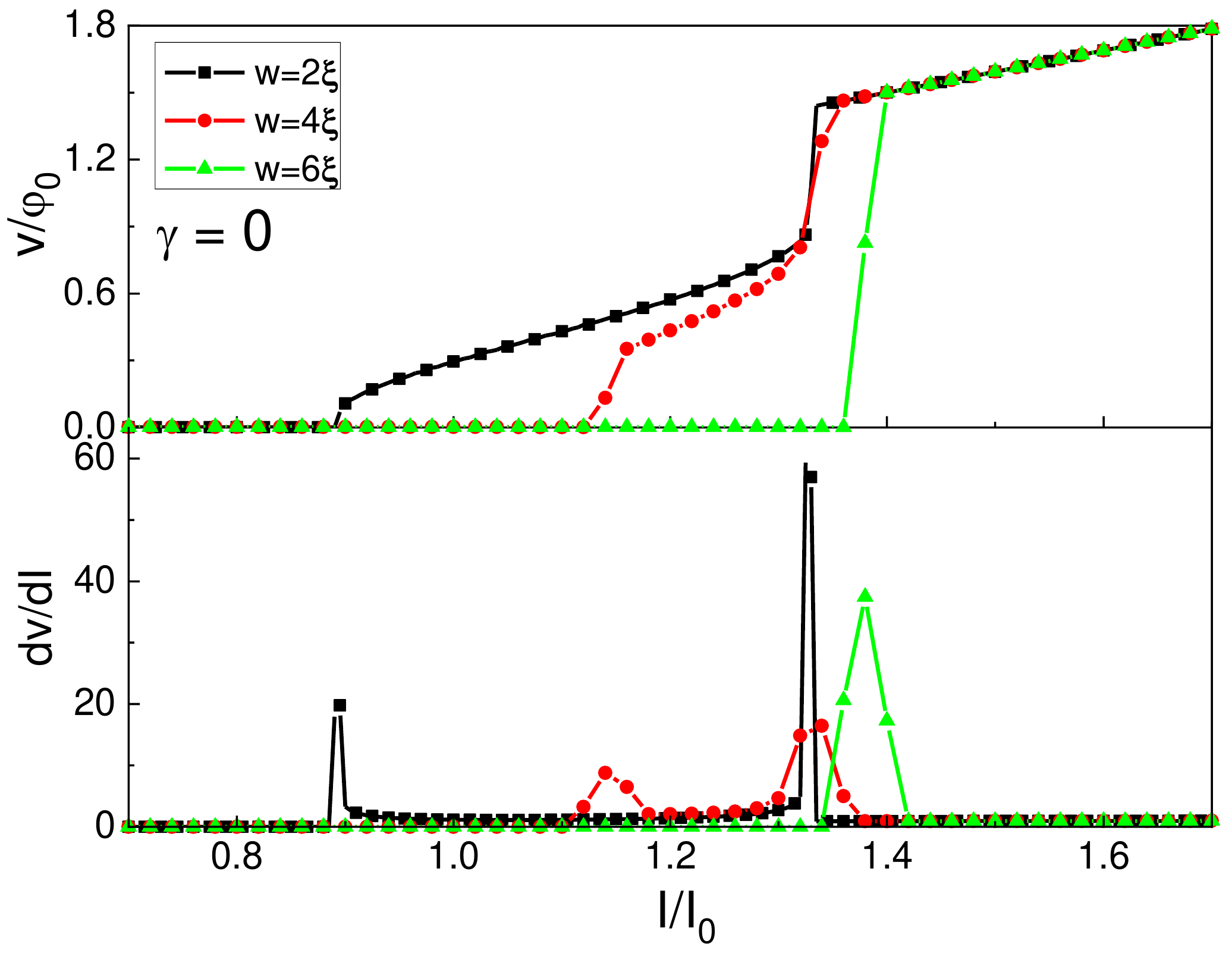}
	\caption{(Color online) $\varphi(J)$ characteristics and differential resistance as a function of the applied current (both normalized) for a gapless system. The size of the constriction was varied by keeping  $\alpha=d=2\xi\left(0\right)$. It is noted that the $J$-range of the resistive state decreases with increasing the constriction size. }
	\label{fig12}
\end{figure}

Figure \ref{fig12} shows the $\varphi(J)$ characteristics and their differential resistance, $d\varphi/dJ (J)$, for gapless systems with different sizes of constrictions, and $\alpha=2\xi(0)$. $J_{c1}$ is $w$-dependent and it is $0.895J_0$, and $1.14J_0$, for $w=2\xi(0)$, and $w=4\xi(0)$, respectively. Meanwhile, the range of currents of the resistive state decreases with $w$, and the current transition to the normal state,$J_{c2}$, occurs at quite the same value for $w=$2 and 4$\xi(0)$. For $w=6\xi(0)$, $J_{c2}$ is larger than for the other samples. In such a case, the constriction is too large that the sample behaves as a homogeneous stripe.

\begin{figure}[!htb]
	\centering
	\includegraphics[width=8cm]{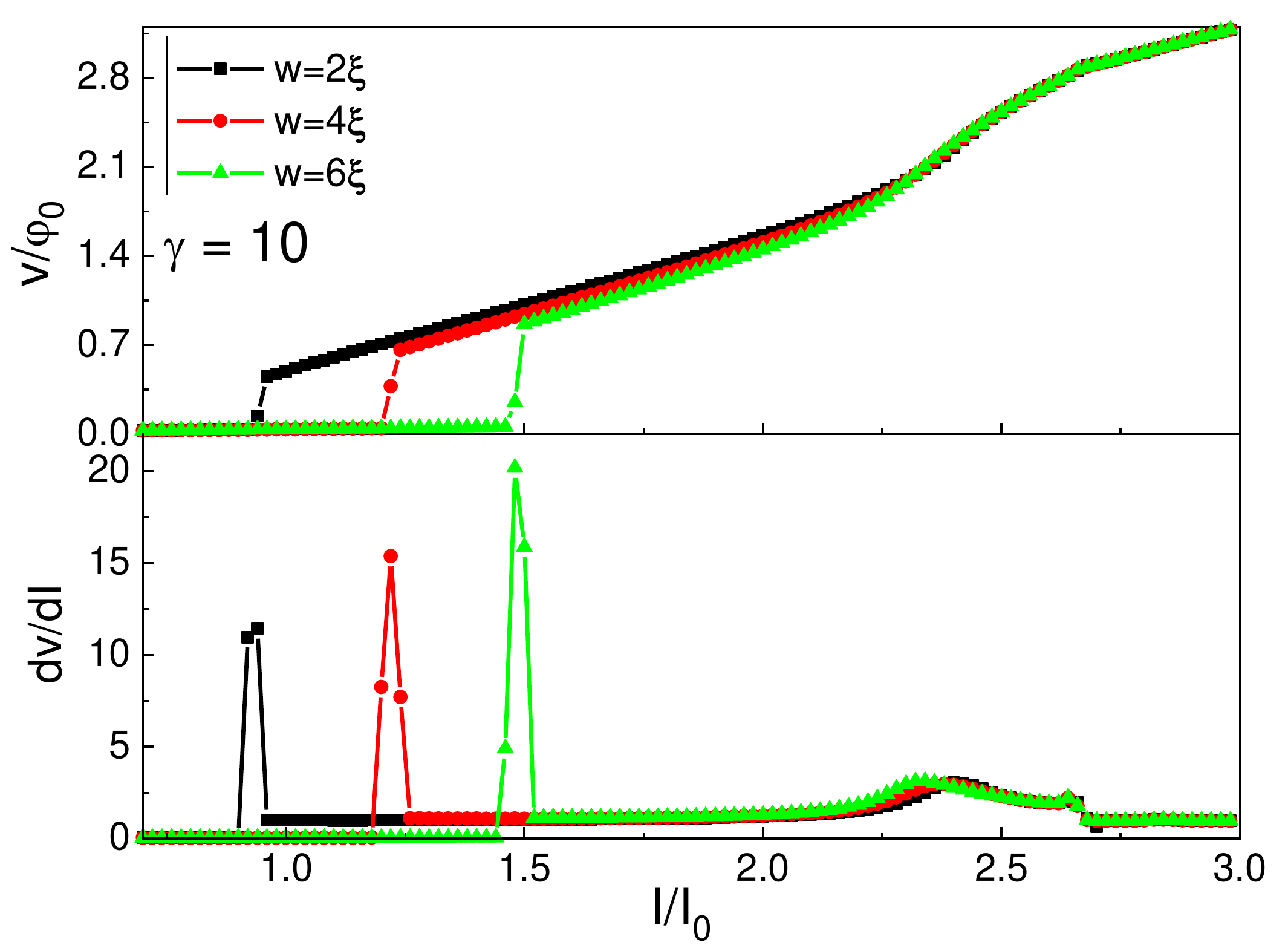}
	\caption{(Color online) $\varphi(J)$ and $d\varphi/dJ(J)$ (both normalized) for a gap-like system. The size of the constrictions was varied by keeping the contact metallic and constriction width size fixed at $\alpha=d=2\xi\left(0\right)$. It is noted that range of resistive state decreases with increasing constriction size.}
	\label{fig13}
\end{figure}

Figure \ref{fig13} shows the $\varphi(J)$ characteristics, and $d\varphi/dJ(J)$, for the gap-like systems. The $J_{c1}$ are $0.92 J_{0}$, $1.20 J_{0}$, and $1.46 J_{0}$ for $w=2\xi(0)$, $w=4\xi(0)$, and $w=6\xi(0)$, respectively. It can also be noticed that the $J$-range of the resistive state decreases by increasing $w$. However, the transition to the normal state occurs at similar current densities, $J_{c2}=2.68J_0$, $2.66J_0$, and $2.66J_0$ for $w=2\xi(0)$, $w=4\xi(0)$, and $w=6\xi(0)$, respectively. 

The gap-like systems also present a region of a partially normal state, as called in Ref. \cite{Beryorovb2009b}. This region appears close to $J_{c2}$ as a second peak in the differential resistance. Such a state is characterized by superconducting regions only at the edges of the samples.

For short, the size of the constriction $w$ controls $J_{c1}$, $J_{c2}$, and, consequently, the range of currents for which resistive states are associated. Besides that, the KV-KAv dynamics remain the same as presented in the previous topic.

\subsection{Voltage-time characteristics}

\begin{figure}[!htb]
	\centering
	\includegraphics[width=8cm]{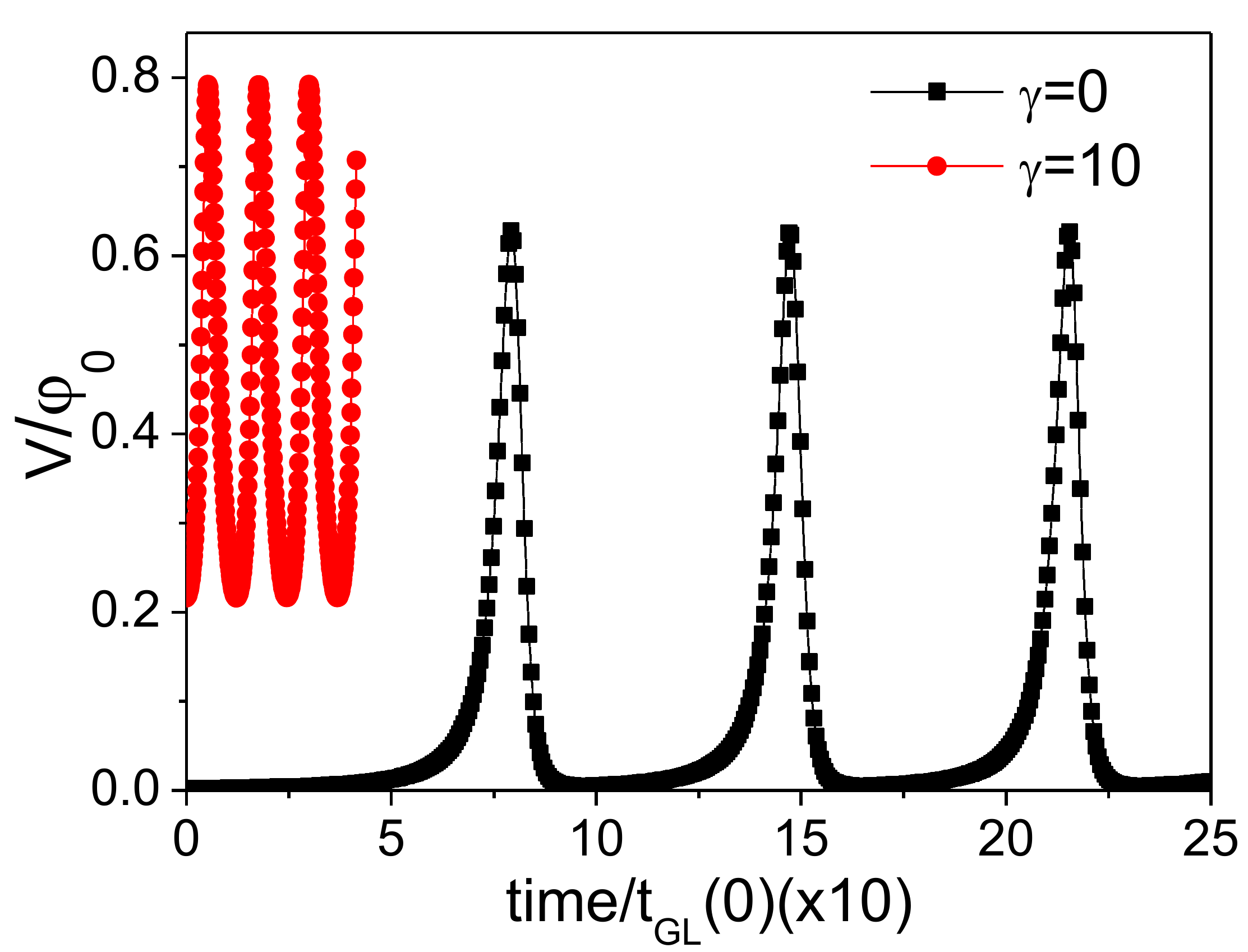}
	\caption{(Color online) Voltage as a function of time for gapless and gap-like superconductors. The curves were obtained for $\alpha=d=w=2\xi(0)$ at $J_{c1}$ of both samples.}
	\label{fig14}
\end{figure}

The temporal oscillations of the voltage produced by the KV-KAv dynamics are affected by the nature of the system. For gap-like samples at $J_{c1}$, $\varphi$ oscillates sinusoidally, and gapless samples present pulse-like oscillations due to a faster $|\psi|$ relaxation ($\gamma=0$), as shown in Figure \ref{fig14}. Besides that, the oscillations reach zero voltage periodically and just after an annihilation event. The non-zero minimum for $\varphi$ is due to the delay in recovering $|\psi|$, as indicated in equation \ref{eq2} \cite{Beryorov2007}, which demands a longer relaxation time.

\begin{equation}
\label{eq2}
\tau_{|\psi|} = {u}\sqrt{1+(\gamma|\psi|)^2}{t}_{GL}(0)
\end{equation}       

\begin{figure}[!htb]
	\centering
	\includegraphics[width=8cm]{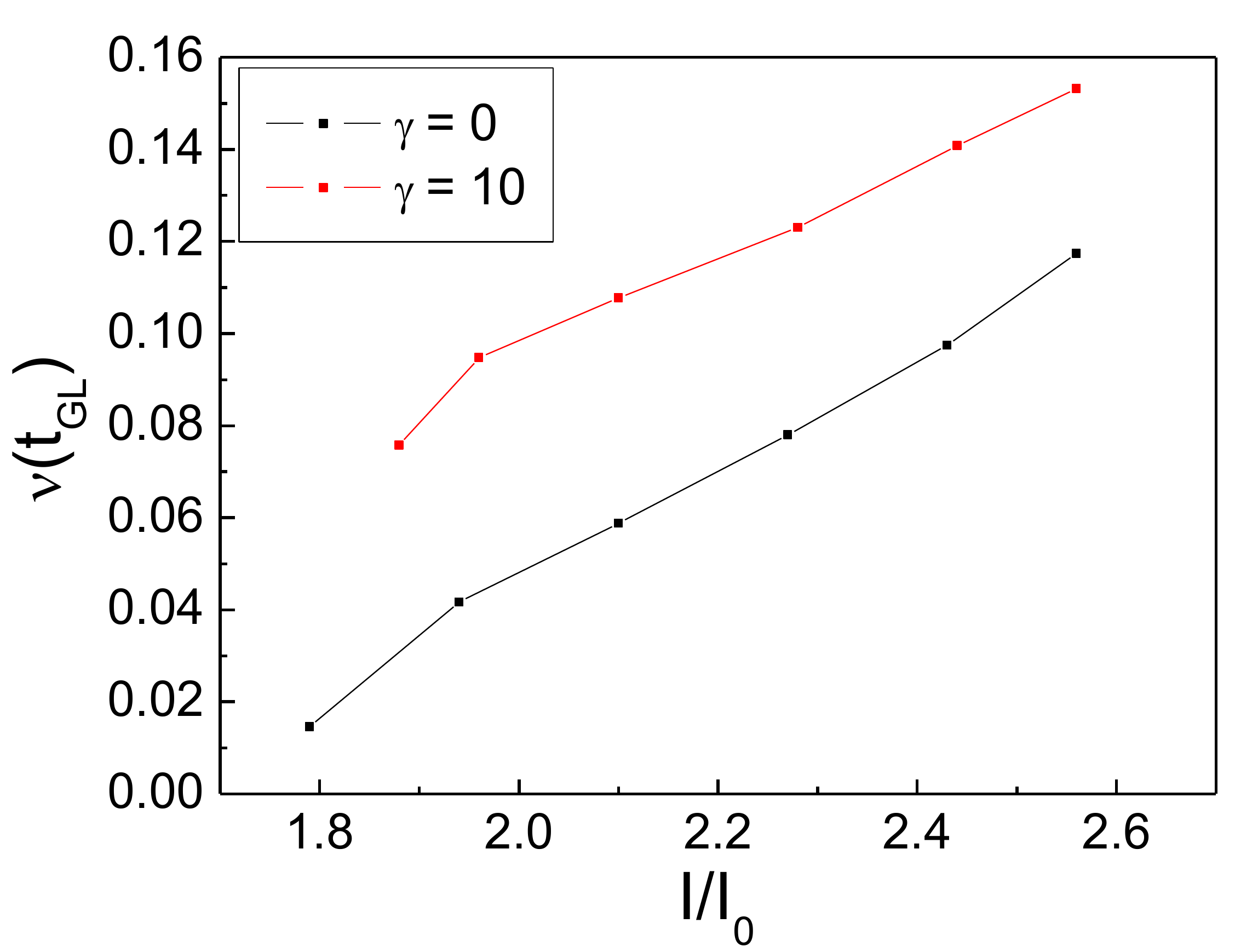}
	\caption{(Color online) Kinematic vortex frequency by applied current for gapless and gap-like systems with $\alpha=w=d=2\xi\left(0\right)$.}
	\label{fig15}
\end{figure}

As expected from the average velocity (Figure \ref{fig11}), the frequency of $\varphi(t)$ is higher for the gap-like sample. Figure \ref{fig15} shows the normalized frequency as a function of the applied current for both samples with $\alpha=w=d=2\xi\left(0\right)$. The relaxation of $\psi$ deteriorates the local superconducting properties, producing an easy way to the vortex flow. Then, the vortex (or antivortex) experiences a lower drag force, reaching high velocities and higher frequencies for annihilation events.
Using the experimental parameters of the PbIn alloy, from the data of Figure \ref{fig15}, the frequencies are of the order of THz, i.e., the range of frequencies for $\gamma =0$ is  (0.135-1.35) THz, and for and $\gamma = 10$ is (0.95-2.00) THz.
As related previously in the literature \cite{Beryorovb2009b}, the KV lifetime is much smaller than the period of ${\varphi}(t)$ oscillations for gap-like samples. In addition, the same behavior was shown by gapless samples. Using the collision time $(t_{KV})$ from Figure \ref{fig7} (square black line) and the period $(t_P)$ of ${\varphi}(t)$ oscillations at ${J}_{c1}$, one obtains ${t}_{KV} = 0.133$ ps and ${t_P} = 7.41$ ps, respectively. However, as pointed out in Ref.~\cite{Beryorovb2009b}, for gap-like samples the rate $\frac{{t_P}}{{t}_{KV}} \propto 1000$. In our approach for gapless samples, $\frac{{t_P}}{{t}_{KV}} \propto 50$, which means that for $\gamma=0$, $t_P > t_{KV}$, i.e., it is still considering a relaxation of $\psi$ due to $u=\tau_{|\psi|}$/$t_{GL}(0)$ (see equation \ref{eq2}).

\section{Conclusion}

This work shows that kinematic vortices appear in gapless superconductors with a constriction made by superconducting defects with lower $T_c$. The dynamics presented by these systems are similar to gap-like ones. However, the relaxation of $|\psi|$ plays an important role in the values of $J_{c1}$ (transition to a resistive state), $J_{c2}$ (transition to the normal state), average velocity, and the frequency of $\varphi(t)$ characteristics. The velocity of a vortex (and the frequency of $\varphi(t)$) in gap-like systems is larger than in gapless ones. On the other hand, the studied systems presented just one type of dynamics, i.e., the kinematic vortex-antivortex pairs are formed on the edges of the surface defects annihilating themselves in the center of the constriction. Such behavior is different from that one reported in Ref.\cite{Beryorovb2009b}.
The gapless superconductors with a constriction open the possibility to have resistive states for higher values of ${\Delta{J}}$, unlike presented in Ref.\cite{KramerBaratoff} (without constriction). The frequency of $\varphi(t)$ is of the order of THz for both gap-like and gapless superconductors. 

In Ref. \cite{KramerRangel}, it was demonstrated that the condition $\gamma > 5.5$ is necessary to occur the phase-slippage in superconductors. On the other hand, we prove the possibility of obtaining a stable resistive phase-slip state for $\gamma < 5.5$ in a superconducting material with surface defects. 
Additionally, from the authors' point of view, this is the first work showing the possibility to have kinematic vortices treating superconducting systems by the gapless time-dependent Ginzburg-Landau theory.


\textbf{Acknowledgements}

We acknowledge the Brazilian agencies S\~ao Paulo Research Foundation (FAPESP, grant 2016/12390-6), Coordena\c c\~ao de Aperfei\c coamento de Pessoal de N\'ivel Superior - Brasil (CAPES) - Finance Code 001, National Council of Scientific and Technological Development (CNPq, grant 302564/2018-7).

\textbf{Author contributions statement}

VSS and ECSD: Validation, Investigation, Data Curation, Writing - Original Draft.
ES: Writing- Reviewing and Editing, Supervision, Validation, Conceptualization, Methodology.
RZ: Resources, Writing- Reviewing and Editing, Visualization, Supervision, Project administration, Funding acquisition, Conceptualization.

\bibliography{referencias}
\end{document}